# The Study State Analysis of Tandem Queue with Blocking and Feedback

**C Chandra Sekhar Reddy**
Asst. Prof., RGM Engg. College, Nandyal, JNTU
E-mail:srisaimax@gmail.com
**K Ramakrishna Prasad**
Professor & BOS, S V University, E-mail:ramk235@gmail.com
**Mamatha**
Asst. Prof., Syamaladevi Engg. College, Nandyal, JNT University

**ABSTRACT**

Computer system models provide detailed answer to system performance. In this paper a two stage tandem network system with Blocking and Feedback is considered and it performance has been analyzed by spectral expansion method. The study state system with balance equations has been discussed.

## 1. Introduction

The structure of a simple two station Markovian tandem queue is as follows: Tasks arrive at a buffer with zero, finite or infinite capacity, in front of the first station according to a Poisson process. Zero capacity means that no task may wait or it will be lost from the system. The tasks are served, in the order of their arrival, by a single server facility with exponential distribution. They then proceed to a buffer at the second station where they are similarly served. Tandem queues with finite buffers have been studied extensively in the past half century, beginning with by Hunt [1]. They are very useful in modeling and analysis of some discrete event systems. In industry, manufacturing flow line systems are modeled as tandem queues with finite buffers, Dallery and Greshwin [2]. A machine is represented as a server and a buffer is represented as storage Dallery and Frein [3].

In classical queuing networks service stations do not exchange information about their queue lengths. However, in general such communication networks may be useful. Suppose for instance that when the queue at some *down-stream* station builds up, this station can protect itself by signaling *upstream* station to decrease or block their service rate. In this way there is congestion dependent feedback information (not for jobs) from downstream station to upstream stations. In addition, tandem networks for two stages, the first server ceases to work when the queue length at the second station hits a *blocking threshold*. After completing processing at second station, a task may leave the system or go back to the first station for re-service with feedback probability. The study state joint distributions of the queue lengths that will lead to the moments of the queue lengths and other performance measures are studied.

When the buffer size is finite in tandem, there is a possibility that the system could become blocked. The study of tandem queues without buffers with blocking was studied in 1965 by Avi-Itzhak and Yadin [4]. Three types of blocking are commonly are considered as: Blocking after service (type 1 blocking): When a task or job wants to go from station α to station $\beta$ and station $\beta$ is full, then station $\alpha$ will be blocked. Now the service that station may stop until a space becomes available for jobs to move onto station $\beta$. Blocking before service (type 2 blocking): When a task $\alpha$ declares its destination, say station $\beta$ prior to starting service and station $\beta$ is full, the server in station $\alpha$ becomes blocked and service stops until a space becomes available in station $\beta$. Repetitive service (type 3 blocking): When a task in station $\alpha$ completes its service and needs to move to station $\beta$ but station $\beta$ is full, so that task $\alpha$ continues to receive service repeatedly until station $\beta$ becomes available.

Comparisons of these types of blocking are discussed in Onvural and Perros [5]. A matrix geometric approximation for tandem queues with blocking and repeated attempts is discussed in Gomez-Corral [6]. Blocking after service in a system that allows priorities and conduct product form approximation based on the principle of maximum entropy for finite capacity open queuing network models is also studied in Kouvtsos and Awan [7]. Performance evaluation of an open queuing network with blocking after service is analyzed. A two station tandem queue with blocking after service that has an exact solution and exact stationary distribution is considered in Akyildiz and Brand [8].

Neuts [9] studied a two station tandem queue with finite intermediate buffer. A general service distribution is assumed for the first station, while in the second station the distribution is exponential. He gave the formula for the time dependent case with emphasis on the stationary process. He also addresses the busy period. Dependent and independent tandem queues with blocking are compared in Browning [1998]. A two station exponential tandem queue with blocking and driving analytic solution for the joint probability distribution in equilibrium using generalized eigenvalues is considered in Grassman and Drekic [10]. Neuts gives the condition for stability of an exponential multi server tandem queue with blocking. He offers an algorithm for obtaining important features of the queue.

A Novel approach to analyze QBD process by spectral expansion method has been intensively developed Chakka [11], further analyzed for approximation [12, 13]. It can be applied to calculate the performance of the system efficiently and gives exact solution for queue length and other parameters at each station. The Poisson arrival rate should not exceed a critical value which depends, in a complicated manner on the service rate.

## 2. Modeling of two node tandem system with server slow down, feedback

We now present the model for the two node tandem network with blocking, write it as a QBD processes, and consider its stability conditions. The system consists of a two service stations with an identical single server in each. The network model, illustrated in Fig. 1 is exponential service system. Tasks arrive from an infinite external source according to a Poisson distribution with arrival rate $\sigma$. Each server has a buffer and service rates at first and second stations are $\mu_1$ and $\mu_2$ respectively. The load on the first and second server is $\rho_1=\sigma/\mu_1$ and $\rho_2=\sigma/\mu_2$, respectively. After service completion at the first station, jobs move on to the second station. Once service is completed there also, jobs leave the network with probability $\theta$ or feed back to first station with probability $1-\theta$.

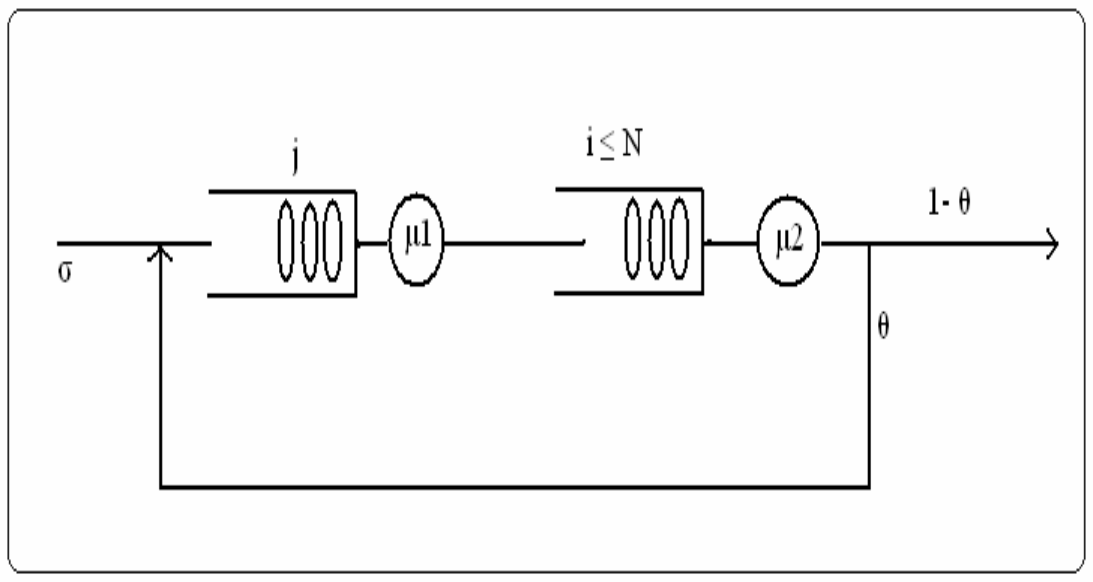

*Figure 1 Two node Tandem queue with feedback, blocking and slow down*

The second station is allowed to inform the first station about the number of jobs in the queue. Immediately after the second station contains $N$ jobs, it signals the first server to stop processing any job in service and servicing jobs remains at first station. In this case server one is said to be blocked. We assume that the feedback signal from the second station to the first station is not delayed. When the queue length in the second station becomes less than $N$, the first server may resume service again. Clearly this blocking mechanism will protect the second station from overflow, at the cost of a stochastically longer queue at the first station.

First, we present the effect on the first station as a function of the blocking threshold $N$. The Markov process system is considered as $X = \{I(t), J(t), t \geq 0\}$, where $I(t)$ represent the number of requests in second buffer, including the one in service. Where as $J(t)$ is the number of requests in first buffer, including the one being served by the server. When $I(t)$ is equal to threshold $N$, the first server blocks., i.e. its service rate becomes zero. Right after the departure of the job in service at the second station, the first server resumes service (if jobs presents there, of course). It is clear that the joint process is a continuous time Markov chain. We assume the Markov process is irreducible and jobs arriving to the system are exponentially distributed.

## 3. The Steady State Solution

The evaluation of the Markov modulated queuing processes is assumed to be governed by three types of transitions. Those are:

- $Aj$ : purely phase transitions – From state $(i, j)$ to state $(k, j)$ $(0 \leq i, k \leq N;\ i \neq k;\ j = 0, 1, \ldots)$
- $Bj$ : one-step upward transitions – From state $(i, j)$ to state $(k, j+1)$ $(0 \leq i, k \leq N;\ j = 0, 1, \ldots)$
- $Cj$ : one-step downward transitions – From state $(i, j)$ to state $(k, j-1)$ $(0 \leq i, k \leq N;\ j = 0, 1, \ldots)$.

It is further assumed that there is threshold M, which is one for this two stage tandem network, such that the transition rates $Aj$, $Bj$ and $Cj$ do not depend on $j$ when $j \geq$ M. thus, although the system parameters may depend in an arbitrary way on the environmental phase, they ceases to depend on the queue length when later is sufficiently large.

In order to write balance equations satisfied by the equilibrium probabilities, $P_{i,j}$, it is convenient to introduce the (N+1)(N+1) matrices having the transition rates $Aj(i,k)$, $Bj(i,k)$ and $Cj(i,k)$. By definition the main diagonal of $Aj$ is zero and $C_0$, these matrices given as,

The phase transition matrices $A$ is given by:

$$A = Subdiag[\mu_2(1-\theta),\ \mu_2(1-\theta),\ \ldots\ldots\mu_2(1-\theta),]$$

$$for \qquad j = 0,1,2,3,\ldots\ldots$$

One step downward transitions happen when a request goes from stage one to stage two, after service in stage one is completed.

The matrices C and $C_j$ are given by:

$$C = Supdiag[\mu_1, \mu_1, ...., \mu_1] \quad \text{for j=1,2,3,..}$$

Similarly the one step upward transitions matrix, when a new job arrives at stage one or a job after having processed by stage two is fed back to stage one for re-service is given in matrix form B as:

$$B = \begin{bmatrix} \sigma & & & & \\ \mu_2\theta & \sigma & & & \\ & \mu_2\theta & \sigma & & \\ & & \text{O} & \text{O} & \\ & & & \mu_2\theta & \sigma \end{bmatrix}$$

Tandem network system, in this paper is considered with one buffer finite and the other infinite. The study state probabilities of the system considered can be expressed as:

$$p_{i,j} = \lim_{t\to\infty} P\{I(t) = i, J(t) = j\}; \quad 0 \le i \le N, \quad 0 \le j \le \infty \qquad (1)$$

It is convenient to define probability vectors corresponding to state with *j* jobs in the system as:

$$v_j = (P_{0,j},\ P_{1,j},\ ....,P_{n,j})\ ; \quad j = 0,1,2,..... \qquad (2)$$

Then the balance equations for the equilibrium probabilities can be written as

$$V_0[D^A + D^B + D^C] = V_0A + V_1\ C \quad \text{when j=0} \qquad (3)$$

Where $D^A$, $D^B$, $D^C$, are diagonal matrices whose $i^{th}$ element is the sum of $i^{th}$ row sum of the matrices A's, B's, C's respectively of size *(N+1)*(*N+1*).

When *j* is greater than the threshold *M*, the equations, for finite buffer (L), it becomes

$$V_j[D^A + D^B + D^C] = V_{j-1}B + V_jA + V_{j+1}C \quad \text{when j=1,2,......L-1}$$

(4)

In addition, all probabilities must sum up to 1:

$$\sum_{j=0}^{\infty} \mathbf{v}_j e = 1.0$$

(5)

Where *e* is a column matrix whose elements are 1of size N+1.

### 4. Computational processes of the QBD system

The balance equations with constant coefficients (5) are usually written in the form of 2 order homogeneous vector difference equation.

$$v_jQ_0 + v_{j+1}Q_1 + v_{j+2}Q_2 = 0, \qquad j \ge M \qquad (7)$$

where $Q_0$ = B, $Q_1 = A - D^A - D^B - D^C$ and $Q_2$ = C. With the Equation (7) which is called 'characteristics matrix polynomial', Q(x), defined as

$$Q(\lambda) = Q_0 + Q_1\lambda + Q_2\lambda^2$$

(8)

Let $\lambda_k$ be the *generalized eigenvalues,* of Q(λ) in the interior of the unit disk, and d be their number. Denote its corresponding *generalized eigenvectors* with $u_k$. These eigenvalues and eigenvectors satisfy the equation:

$$\det[Q(\lambda_k)] = 0\ , |\ \lambda_k| < 1,\ k = 1,2,...d, \qquad (9)$$

det[Q($\lambda_k$)] means the determinant of Q($\lambda_k$). Also from properties matrix algebra

$$u_k\, Q(\lambda_k) = 0,\ k = 1,2,...\ d. \qquad (10)$$

Assume that the above eigenvalues are simple. This assumption has been satisfied in all examples that have been examined numerically; in some cases it has been proved analytically. In fact, a weaker assumption would suffice, namely that if an eigenvalue has multiplicity m, then it also has m linearly independent left eigenvectors.

The following result provides the 'spectral expansion' solution of the Markov-modulated queue.

**Proposition:** The QBD process is study state, if and only if, the number of eigenvalues of Q(λ) in the interior of the unit disk is equal to the number of states of the Markovian environment, i.e. d = N + 1. Then the solution of (7) has the form:

$$v_j = \sum_{k=1}^{N+1} \alpha_k u_k \lambda_k^j, \qquad j \ge M$$

(11)

where $\alpha_k$, k = 1, 2, … N+1 are some constants may be complex too

Note that if there are non-real eigenvalues in the unit disk, then they appear in complex conjugate pairs and its corresponding eigenvectors are also complex conjugates. From this it must be true that the appropriate pairs of complex constants $\alpha_k$, in order that the right hand side of (11) be real.

From the balance equation (4), the constant coefficients $\alpha_k$ and probability vectors $v_j$ for j = 0, 1, 2, … M are to be determined. This is a set of (M+1)(N+1) linear equations with M(N+1) unknown probabilities and the N+1 constants $\alpha_k$. Since the generator matrix is singular hence only (M+1)(N+1)–1 of these equations are linearly independent. To make the generator matrix non singular it requires another equation and this job will be done by (6).

The quadratic eigenvalue-eigenvector problem (10), for computational purposes can be reduced to a linear form uQ = λu, where Q is a matrix of size (2N+2)(2N+2). The process of valuation, in detail discussed R Chakka in [10].

In what follows, Proposition 1 will be used to derive approximations, rather than exact solutions. A central role in those developments is played by the largest eigenvalue, $\lambda_{N+1}$, and its corresponding left eigenvector. When the queue is stable, $\lambda_{N+1}$, is real, positive and simple.

Moreover, it has a positive eigenvector. From this onward, $\lambda_{N+1}$, will be referred to as the *dominant eigenvalue*, and is denoted by γ.

Expression (11) implies that the tail of the joint distribution of the queue size and the environmental phase is approximately geometrically distributed, with parameter equal to the dominant eigenvalue, γ. To see that, divide both sides of (11) by $\gamma^{j}$ and j→∞. Since γ is strictly greater than in modulus than all other eigenvalues, all terms in the summation vanish, except one:

$$\underset{j\to\infty}{Lt}\frac{v_j}{\gamma^j}=\alpha_{N+1}u_{N+1} \tag{12}$$

When j is large, the above form can be expressed as:

$$v_j \approx \alpha_{N+1}u_{N+1}\gamma^j \tag{13}$$

This product form implies that when the queue is large, its size is approximately independent of the environment phase. The tail of the marginal distribution of the queue size is approximately geometric:

$$p_{.,j} \approx \alpha_{N+1}(u_{N+1}.e)\gamma^j \tag{14}$$

Where ***e*** is the column matrix defined earlier.

These results suggest seeking an approximation form:

$$v_j = \alpha u_{N+1}\gamma^j \quad \text{where } \alpha \text{ is a constant.} \tag{15}$$

Note that γ and $u_{N+1}$ can be computed without having to find all eigenvalues and eigenvectors. There are techniques for determining the eigenvalues that are near a given number. Here we are dealing with the eigenvalue that is nearest to but strictly less than 1.

One could decide to use the approximation (15) only for $j \geq M$. Then the coefficient $\alpha$ and the probability vectors $\mathbf{v}_j$ for $j = 0, 1, \ldots, M-1$ can be obtained from the balance equations (4), e.g., for $j < M$, and the normalizing equation (6). In that case, one would have to solve a set of $M(N+1)+1$ simultaneous linear equations with $M(N+1)+1$ unknowns. If that approach is adopted, then the approximate solution satisfies balance equations of the Markov process except those for $j = M$.

Alternatively, and even more simply, (15) can be applied to all $\mathbf{v}_j$, for $j = 0, 1, \ldots$ Then the approximation depends on just one unknown constant, $\alpha$. Its value is determined by (6) alone, and the expressions for $\mathbf{v}_j$ become

$$v_j = \frac{u_{N+1}}{(u_{N+1}.e)}(1-\gamma)\gamma^j,\ j=0,1,2,\ldots \tag{16}$$

and the mean queue length is calculated by

$$MQL = \frac{u_{N+1}}{(u_{N+1}.e)}\frac{\gamma}{(1-\gamma)}, \tag{17}$$

This last approximation avoids completely the need to solve a set of linear equations. Hence, it also avoids all problems associated with ill-conditioned matrices. Moreover, it scales well. The complexity of computing γ and $u_{N+1}$ grows roughly linearly with $N$ when the matrices $A$, $B$, and $C$ are sparse. The price paid for that convenience is that the balance equations for $j \leq M$ are no longer satisfied. The geometric approximation is asymptotically exact when the offered load increases i.e. the arrivals or services of the jobs are done rapidly so that system becomes heavily loaded and approaches saturation.

## 5. Numerical Results of the system

Values of parameters in the numerical examples are chosen to satisfy the condition of existence of stationary processes and the stability conditions. Mat Lab software is used to obtain the accompanying graphs for the numerical examples. The proposed work to present the numerical results, if not mentioned, the parameters is considered as N=10, $\mu_1 = 3.0$, $\mu_2 = 2.5$ and feed back probability $\theta = 0$.

In tandem system two processors are arranged in sequential order, each system has independent queue. The queue which is between the servers is buffer queue with finite length and the other queue is considered either finite or infinite length. Figure 3 is represented for both finite and infinite length where as all remaining results are studied for infinite length. Fig. 3 is drawn between mean queue length and jobs arrival rate with various queue length. It is observed that as queue length decreases mean queue length also decreases. It is due to rejecting pockets for finite queues.

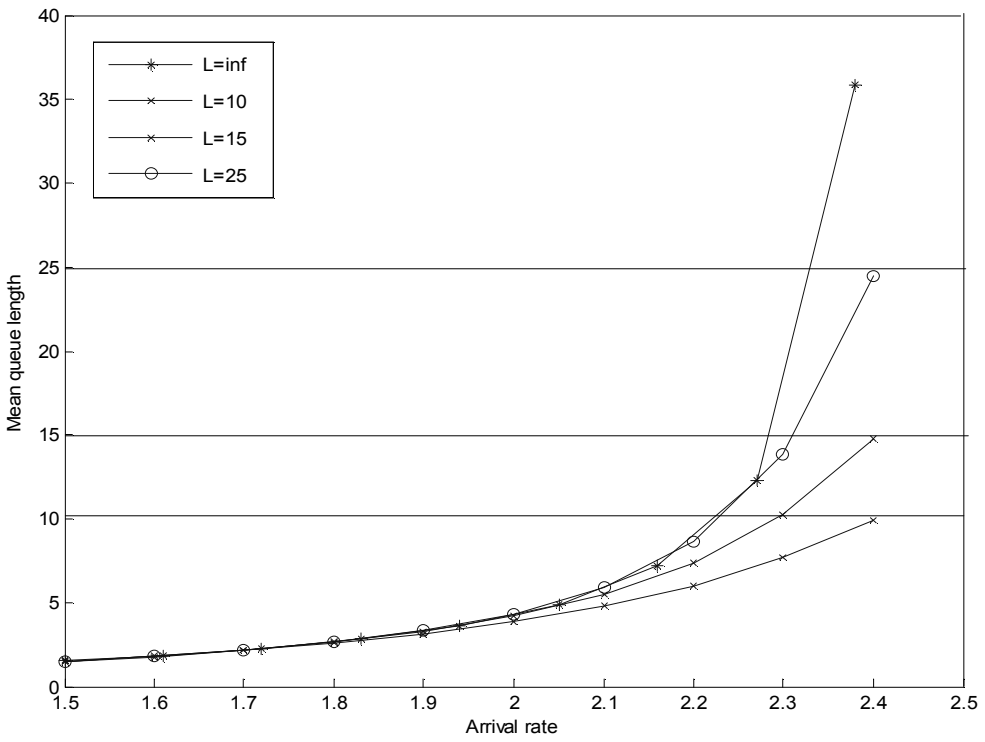

Fig. 3 Mean queue length vs. arrival rate with blocking

From fig. 4 it is observed that as feedback probability increases then mean queue length also increases. This graph has been drawn for various constant arrival rates. It is observed from fig.5 that as buffer size increases, for constant arrival rate of jobs, mean queue length decreases and reaches to constant level after some particular length of the buffer. Fig. 6 is presented for mean queue length arrival rate of fixed feed back probability and observed that as arrival increases then mean queue length increases and as feed back increases mean queue length increases. Where as fig. 7 is drawn for mean queue length with buffer size for various feed back probability, it is observed mean queue length decreases as buffer size increases and mean queue length increases for feed back increases. Here arrival rate σ = 1.5 has been considered.

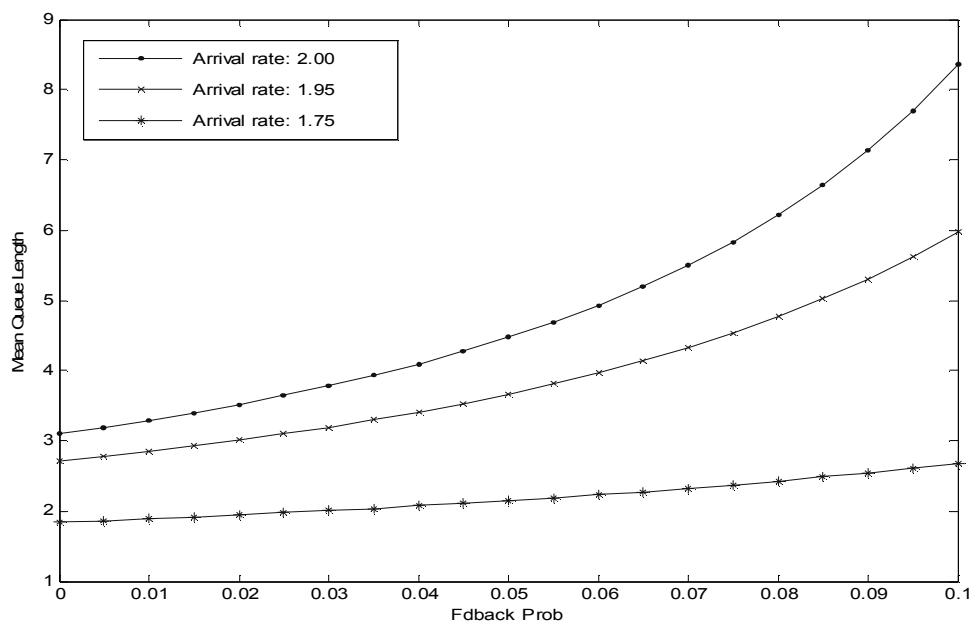

Fig. 4 Mean queue length vs. $\theta$ for various arrival rates

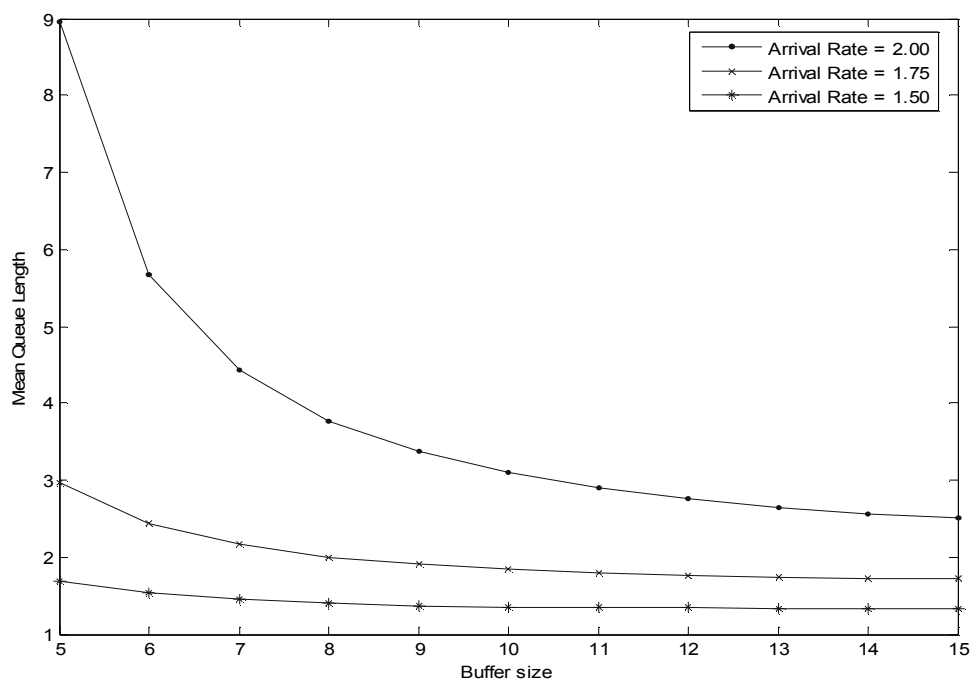

Fig. 5 Mean queue length vs. Buffer size

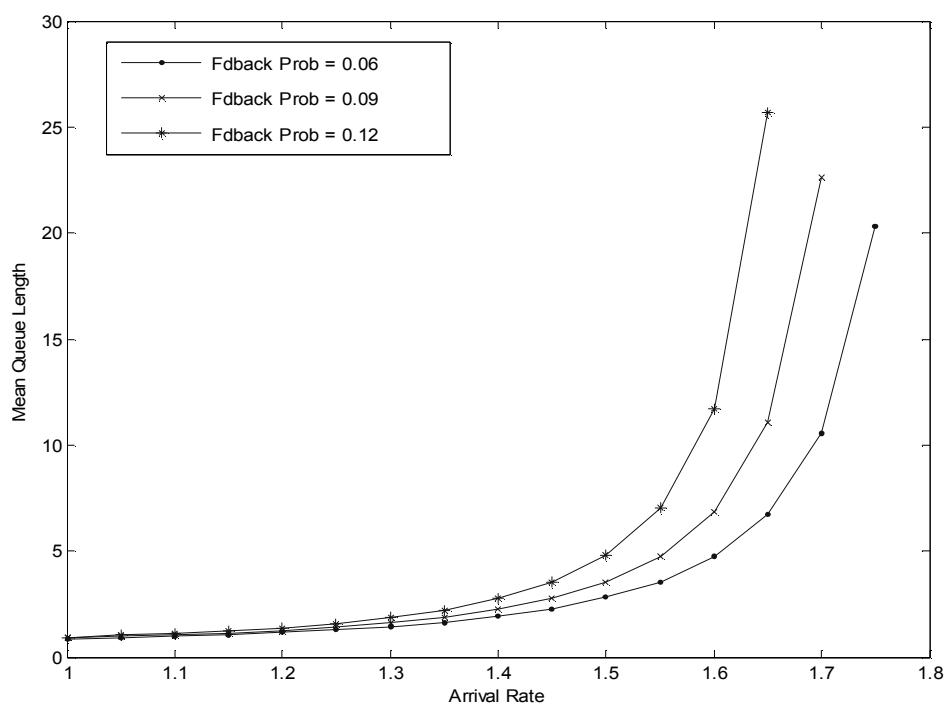

Fig. 6 Mean queue length vs. arrival rate

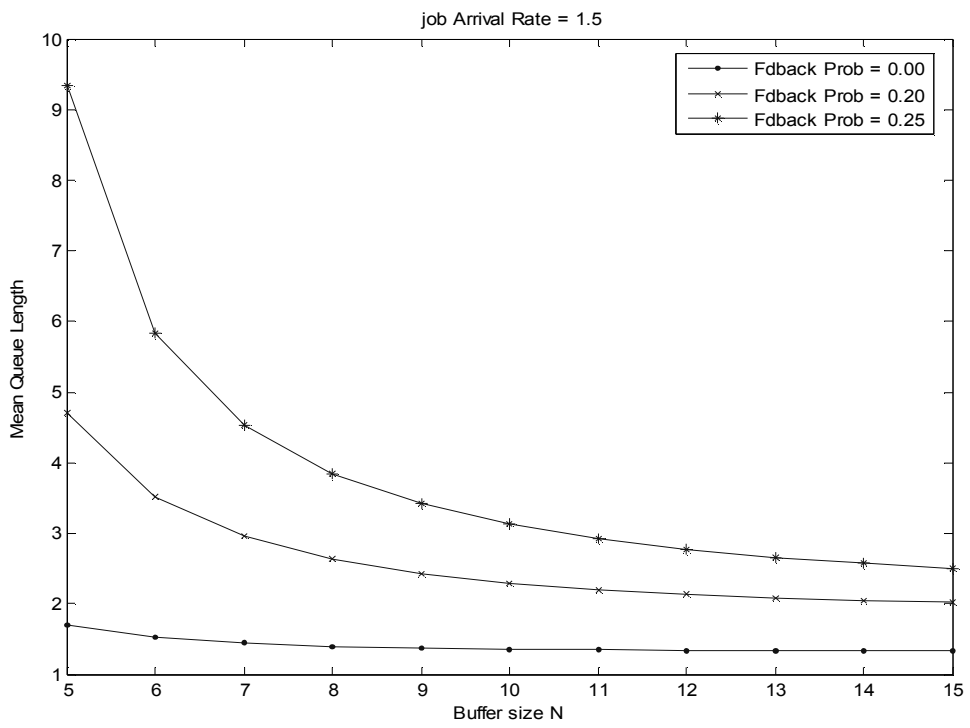

Fig. 7 Mean queue length vs. Buffer size

## 5. Conclusions

In this paper a queuing system with two buffers, one is finite and the other is infinite/finite buffer is considered. The novelty of the model lies in the combination of the features of blocking, feedback and server slowdown. Spectral expansion is considered for the stationary process to determine the distribution of the queue length and other queue parameters as consequences. The numerical examples demonstrate the behavior of the models and some imbalances that occur.

**Authors Biography**

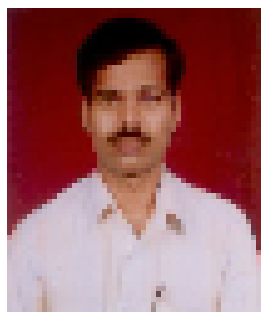

*Mr. Chandra Sekhar Reddy* is pursuing Ph. D. at Sri Venkateswara University, Tirupati. He has presented and published ten technical papers in international conferences and journals. He has been working as Asst. Professor for past ninet years in RGM engineering college, Nandyal. His current areas of research are modeling and performance evaluation of multiprocessor system, Markov processes.

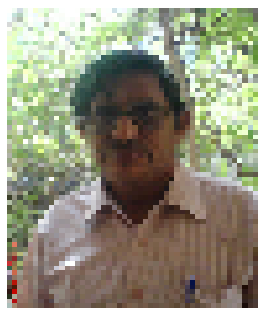

*Dr. K. Rama Krishna Prasad* received his Ph.D. degree from IIT Khanpur. He is having more than three decades of experience in teaching at engineering colleges and P. G. college of S. V. University. Now he is working as senior professor at the Dept. of Mathematics and Board of Studies in S. V. University, Tirupati. He has guided and guiding several Ph. D. students.

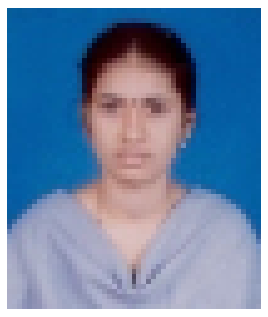

*Mrs. Mamatha* completed her post graduation from Sri Krishnadevara University and presently working as Assistant Professor at Syamaladevi Institute of Science and Technology, Nandyal, Andhra Pradesh. Her research fields include Markov Chains, queuing process, mathematical modeling and fuzzy logic